# Spatial Analysis of Cities Using Renyi Entropy and Fractal Parameters

Yanguang Chen; Jian Feng

(Department of Geography, College of Urban and Environmental Sciences, Peking University, Beijing 100871, P.R. China. E-mail: chenyg@pku.edu.cn; fengjian@pku.edu.cn)

**Abstract**: The spatial distributions of cities fall into two groups: one is the simple distribution with characteristic scale (e.g. exponential distribution), and the other is the complex distribution without characteristic scale (e.g. power-law distribution). The latter belongs to scale-free distributions, which can be modeled with fractal geometry. However, fractal dimension is not suitable for the former distribution. In contrast, spatial entropy can be used to measure any types of urban distributions. This paper is devoted to generalizing multifractal parameters by means of dual relation between Euclidean and fractal geometries. The main method is mathematical derivation and empirical analysis, and the theoretical foundation is the discovery that the normalized fractal dimension is equal to the normalized entropy. Based on this finding, a set of useful spatial indexes termed dummy multifractal parameters are defined for geographical analysis. These indexes can be employed to describe both the simple distributions and complex distributions. The dummy multifractal indexes are applied to the population density distribution of Hangzhou city, China. The calculation results reveal the feature of spatio-temporal evolution of Hangzhou's urban morphology. This study indicates that fractal dimension and spatial entropy can be combined to produce a new methodology for spatial analysis of city development.

## 1. Introduction

Cities and systems of cities are complex systems, but a complex system has its simple aspects. In order to understand city development or urban evolution, we must explore the spatial distribution (e.g., urban density distribution) and the related probability distribution (e.g., rank-size distribution) of cities. The spatial distributions and probability distributions can be divided into two types: one is the simple distribution with characteristic scales such as exponential distribution, and the other is complex distribution without characteristic scales such as power-law distribution (Barabasi and Bonabeau, 2003; Chen, 2015; Takayasu, 1990). The complex distribution without characteristic scales can be termed scale-free distribution or scaling distribution. One of powerful tools for scaling analysis is fractal geometry (Mandelbrot, 1982). Fractal theory has been applied to urban studies for a long time (Batty and Longley, 1994; Frankhauser, 1994; Portugali, 2000). The fractal city studies lead to new urban theory. However, the target of complexity science is not for complexity itself, but for the inherent relationships between complex phenomena and simple rules. Therefore, in urban geographical studies, we should explore both the simple and complex aspects and the connection between spatial complexity and simplicity. The limitation of monofractal method is that it is not suitable for simple distributions. The dimension of a non-fractal distribution is a Euclidean dimension and provides us with no geographical spatial information.

However, it is possible to generalize multifractal measures to describe the spatial distributions with characteristic scale. The new indexes are not real multifractal parameters, but bear analogy with multifractal parameters. The main reasons are as follows. First, multifractal parameters are based on entropy functions, and entropy can be employed to measure both fractal and non-fractal distributions. Second, multifractal measures take on several parameter spectrums, and the spectrums compose both fractal parameters and non-fractal parameters. Third, multifractal dimensions are mainly generalized fractal dimension, and Euclidean dimension can be treated as special cases of fractal dimension. If we find a parameter link between simple distributions and complex distributions, we will be able to generalize multifractal theory and apply it to varied spatial distributions. The link rests with the concept of normalized entropy. In fact, all the spatial distribution data can be converted into probability distribution data, and based on a probability distribution, entropy can be evaluated. Entropy is measure of complexity (Bar-Yam, 2004a; Bar-Yam, 2004b; Cramer, 1993). Hausdorff dimension proved to be equivalent to Shannon entropy and Kolmogorov complexity (Ryabko, 1986). The breakthrough point of developing fractal measures is entropy functions and the association of entropy with fractal dimension. A recent finding is that the normalized fractal dimension is theoretically equal to the normalized entropy under certain conditions (Chen, 2012; Chen, 2017). Thus, the key to solving the problem is to establish a methodological framework for measurements of entropy and fractal dimension.

The methodological framework can be defined by analogy with the box-counting method in fractal theory. Thus, spatial entropy can be naturally associated with fractal dimension. The concept and measurement of spatial Shannon entropy has been introduced to geographical analysis for many years (Batty, 1974; Batty, 1976). More general spatial entropy is the spatial Renyi entropy (Chen and Wang, 2013; Fan *et al*, 2017). Both Renyi entropy and fractal dimension can be employed to measure urban sprawl (Padmanaban *et al*, 2017; Terzi and Kaya, 2011), and the similar functions suggest the intrinsic relation between entropy and fractal dimension. The traditional spatial entropy is quantified by means of geographical systems of zones. A zonal system always takes on an irregular network. Based on the irregular nets, entropy cannot be converted into fractal dimension. However, if we use regular grid to replace the irregular network, we will be able to calculate both spatial entropy and fractal dimension. The functional box-counting method is based on the recursive process of regular grid. This method is proposed by Lovejoy *et al* (1987) and consolidated by Chen (1995), and can be applied to both entropy and fractal dimension measurements. Based on the functional box-counting method, the normalized fractal dimension proved to equal the normalized entropy (Chen, 2017). Based on this equivalence relationship, we can extend the multifractals parameters and obtain useful spatial indexes. This paper is devoted to developing the methodology of spatial analysis using spatial entropy and fractal measures. The functional boxes will be replaced by systems of concentric rings. The rest parts are organized as follows. In Section 2, based on the relationships between spatial Renyi entropy and general correlation dimension, the multifractal measures are generalized to yield a set of new spatial measurements for urban studies; In Section 3, the generalized multifractal parameters are applied to the city of Hangzhou, China, to make a case study; In Section 4, the related questions are discussed, and finally the work is concluded by summarizing the main points. The methodology developed in this article may be used to characterize the spatial structure of other natural and social systems.

# 2. Models

## 2.1 Spatial Renyi entropy

It is necessary to clarify the internal relation between spatial entropy and fractal dimension of cities. Both entropy and fractal dimension can serve for the space-filling measures of city development. A central region of a city has higher fractal dimension values (Feng and Chen, 2010), and accordingly, it entropy value is higher than the periphery region (Fan *et al*, 2017). The common fractal dimension formulae are all based on entropy functions. The generalized correlation dimension of multifractals is defined on the base of Renyi's entropy (Renyi, 1961), which is formulated as follows

$$M_q = -\frac{1}{q-1}\ln\sum_{i=1}^{N}P_i^q , \tag{1}$$

where $q$ denotes the moment order, $M_q$ refers to Renyi's entropy of order $q$, $P_i$ refers to the occurrence probability of the $i$th fractal copy, and $N$ to the number of fractal copies ($i$=1, 2, 3,…, $N$). A fractal copy can be treated as a fractal unit in the fractal set. Thus the generalized correlation dimension can be given as (Feder, 1988; Harte, 2001; Mandelbrot, 1999; Vicsek, 1989)

$$D_q = -\frac{M_q(\varepsilon)}{\ln\varepsilon} = \frac{1}{q-1}\frac{\ln\sum_{i=1}^{N(\varepsilon)}P_i(\varepsilon)^q}{\ln\varepsilon} , \tag{2}$$

in which $\varepsilon$ denotes the linear size of fractal copies at given level, and $N(\varepsilon)$ and $P_i(\varepsilon)$ refers to the corresponding fractal copy number and occurrence probability. If we use a box-counting method to make spatial measurements, then $\varepsilon$ represents the linear size of boxes, $N(\varepsilon)$ refers to the number of nonempty boxes, and $P_i(\varepsilon)$ to the proportion of geometric objects in the $i$th box. Based on functional box-counting method, the normalized fractal dimension $D_q^*$ proved to be equal to the normalized entropy $M_q^*$ (Chen, 2012), that is

$$D_q^* = \frac{D_q - D_{\min}}{D_{\max} - D_{\min}} = \frac{M_q - M_{\min}}{M_{\max} - M_{\min}} = M_q^*, \tag{3}$$

where $D_{\max}$ refers to the maximum fractal dimension, $D_{\min}$ to the minimum fractal dimension, $M_{\max}$ refers to the maximum entropy, and $M_{\min}$ to the minimum entropy. In theory, for the fractal cities defined in a 2-dimensional embedding space (Batty and Longley, 1994), the basic parameters are as follows: $D_{\max}$=2, $D_{\min}$=0, $M_{\max}$=ln$N_T$, $M_{\min}$=0, and $N_T$ is total number of all possible fractal units or boxes (nonempty boxes and empty boxes) in a study area. Thus, equation (3) can be reduced to

$$D_q^* = \frac{D_q}{D_{\max}} = \frac{M_q}{M_{\max}} = M_q^*, \tag{4}$$

which is valid only for fractal systems. This suggests that the ratio of the actual fractal dimension to the maximum fractal dimension is theoretically equal to the ratio of the actual entropy to the maximum entropy of a fractal object. The relation between entropy and fractal dimension is supported by the observational data of cities such as Beijing and Hangzhou (Chen, 2017).

## 2.2 Generalized multifractal parameters

Fractal dimension can be only applied to fractal objectives, and we cannot use fractal geometry to model the non-fractal phenomena. In fact, a non-fractal system can be described with the common measures such as length, area, volume, and density rather than fractal dimension. In other words, the standard fractal parameters have strict sphere of application: irregular and self-similar patterns, nonlinear and recursive process, and complex and scale-free distributions (Chen, 1995). In contrast, entropy can be utilized to measure any distributions of cities, including fractals and non-fractals. However, for the fractal distributions, entropy values depend on spatial scales of measurements. Fortunately, using equation (4), we can generalize fractal measurements, and define a set of quasi-fractal parameters for spatial analysis. In fact, for non-fractals, we have $M_q^*= M_q/M_{\max}$, but we have no $D_q^*= D_q/D_{\max}$ because $D_q$ is not existent. Now, we can define a dummy normalized fractal parameter as follows

$$D_q^* = M_q^* = \frac{M_q}{M_{\max}}, \tag{5}$$

in which $M_q$ and $M_{\max}$ are measurable. For a non-fractal object, the maximum dimension is just the Euclidean dimension of the embedding space, that is $D_{\max}=d$. Thus we can further define a dummy multifractal dimension as below:

$$D_q = D_{\max} D_q^* = dM_q^* = \frac{2M_q}{M_{\max}}, \tag{6}$$

which can be theoretically demonstrated (Chen, 2017). If a city is examined in a 2-dimensional space, then $D_{\max}=d=2$. In fractal theory, the mass exponent can be given by

$$\tau_q = (q-1)D_q, \tag{7}$$

in which $\tau_q$ denotes the common mass exponent. Accordingly, we can define a generalized mass exponent such as

$$\tau_q = (q-1)dM_q^* = \frac{2(q-1)M_q}{M_{\max}}, \tag{8}$$

which represents a dummy mass exponent ($d$=2). It is not a scaling exponent, but it is a useful index for urban spatial analysis. Both $D_q$ and $\tau_q$ are global parameters of spatial analysis. Going a further step, we can introduce a pair of local parameters by Legendre transform (Feder, 1988; Stanley and Meakin, 1988)

$$\alpha(q) = \frac{\mathrm{d}\tau(q)}{\mathrm{d}q} = D_q + (q-1)\frac{\mathrm{d}D_q}{\mathrm{d}q}, \tag{9}$$

$$f(\alpha) = q\alpha(q) - \tau(q) = q\alpha(q) - (q-1)D_q, \tag{10}$$

in which $\alpha(q)$ refers to dummy singularity exponent, and $f(\alpha)$ to the dummy local fractal dimension, d denotes derivative. Of course, in this work, $\alpha(q)$ is not a real scaling exponent, and $f(\alpha)$ is not a real fractal dimension. The word "dummy" means "a thing that seems to be real but is only a copy of the real thing", and dummy multifractal measures are not real multifractal measures. However, both $\alpha(q)$ and $f(\alpha)$ are measurable and useful indexes for urban studies, bearing analogy in form with the real singularity exponent and local fractal dimension in multifractal theory. Actually, the generalized multifractal parameters, including the global parameters and local parameters, are not fractal parameters in the sense of scaling. However, they are really multi-scale indexes of spatial distributions of cities, which bear an analogy with real multifractal parameters and have practical function in spatial analysis (Table 1).

**Table 1 The spatial meaning of the indexes based on generalized multifractal parameter**

| Level | Parameter | Symbol | Meaning | Measure |
|---|---|---|---|---|
| **Global level** | Generalized global dimension | $D_q$ | Scale-free spatial entropy | Global space filling |
| | Generalized mass exponent | $\tau_q$ | Product of generalized dimension and order of moment | Global differentiation |
| **Local level** | Generalized singularity exponent | $\alpha(q)$ | Change rate of mass exponent against order of moment | Local variation |
| | Generalized local dimension | $f(\alpha)$ | Local scale-free spatial entropy | Local space filling |

**Note**: Despite the fact that the generalized multifractal parameters are not real fractal parameters, we using the same notation as that in multifractal theory so that it is easy for readers to know the correspondence relation between the indexes proposed in this paper and the parameters for multifractal measures.

# 3. Empirical analysis

## 3.1 Study area, datasets, and method

The city of Hangzhou, the capital of Zhejiang Province, China, can serve as an example to make empirical analyses using the generalized multifractal parameters based on the spatial Renyi entropy. We have four years of population census data for Hangzhou city, and these census data have been processed and converted into spatial datasets of population density distribution (Feng, 2002). In China, the census tracts of cities are termed *jiedao*. The area of *jiedao* bears an analogy with the UK enumeration districts (Longley, 1999), or the US sub-districts (Wang and Zhou, 1999). All the regions of *jiedao* within a metropolitan area compose a system of zones. Because the average area of *jiedao* is significantly larger than the average area of normal census tracts, we have to transform the zonal data into ring data by the ideas from averaged treatment. Taking the core of urban population distribution as the center of circle, we can draw a series of concentric circles. The interval distance of two circles is 0.6 km, and the maximum radius is 15.3 km. Thus a ring comes between two adjacent circles. A center point and 26 circles result in 26 rings ($N$=26) (Figure 1). By spatial weighed average, we can generate the average population density of each ring. The approach and results of data processing have been illustrated by Feng (2002) and discussed by Chen (2008) as well as Chen and Feng (2012). In this work, we directly make use of the spatial datasets to explore the spatio-temporal of Hangzhou cities.

To evaluate spatial Renyi entropy, we can transform the population density distribution into probability density distribution. The formula is as below:

$$P_i = \frac{\rho_i}{\sum_{i=1}^{26}\rho_i} = \frac{\rho(r)}{\sum_{r=0.3}^{15.3}\rho(r)} = P(r), \quad (11)$$

where $\rho_i$ denotes average population density of the $i$ ring ($i$=1,2,…,26), and $P_i$ represents the probability density of the $i$th ring. The number $i$ corresponds to the radius from the population center $r$ ($r$=0.3, 0.9, …, 15.3). The results are shown in Table 2. Then, using equation (1), we can calculate the Renyi entropy. According to Hospital's rule, if $q$=1, Renyi entropy will change to Shannon's information entropy (Shannon, 1948), and equation (1) should be replaced by

$$M_1 = H = -\sum_{i=1}^{N} P_i \ln P_i \,, \tag{12}$$

where $H$ denotes information entropy. Shannon's information entropy has similar measurement function to Renyi's entropy in urban studies (Fan *et al*, 2017). In fact, Shannon entropy has been applied to geographical spatial analysis many years ago (Batty and Sammons, 1979). The maximum entropy value is $M_{\max}$=ln$N$= ln(26). By means of equations (6) and (8), we can compute the dummy correlation dimension $D_q$ and mass exponent $\tau_q$.

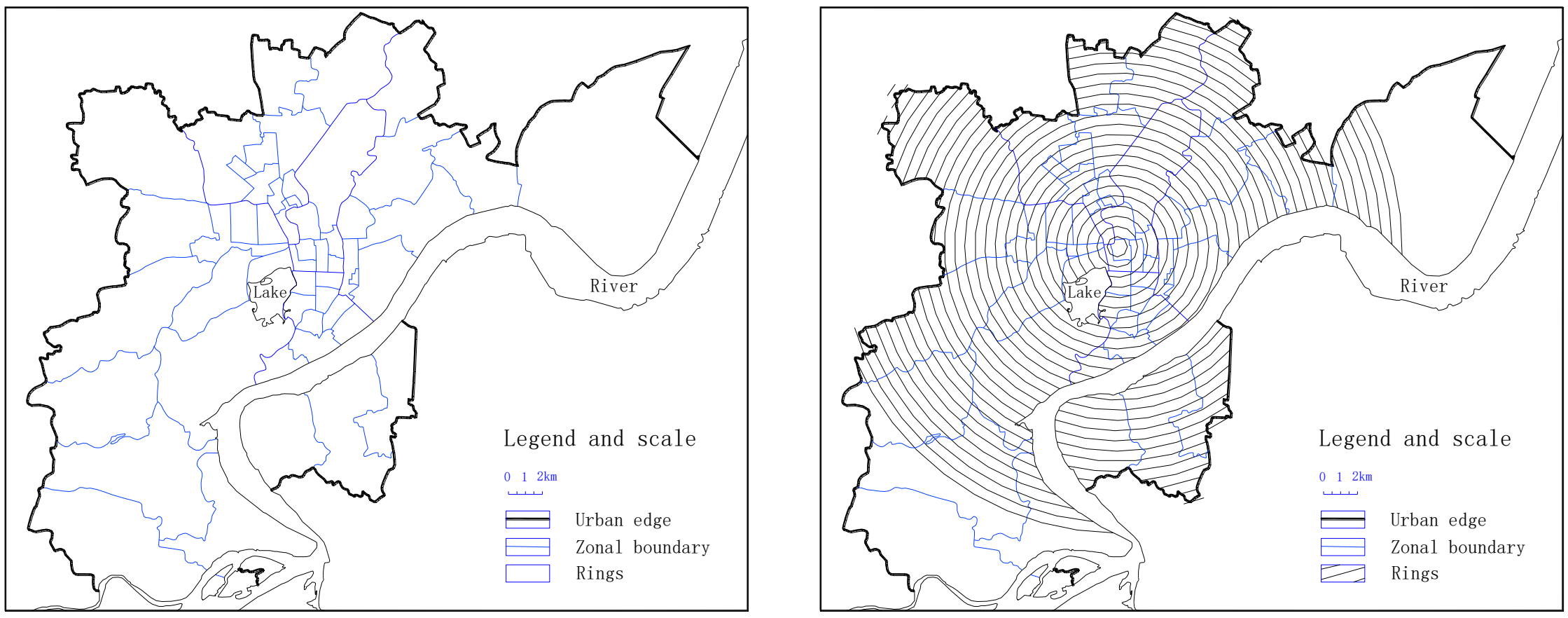

a. Zonal system b. Concentric rings

**Figure 1 The zonal system of census tracts and concentric rings in the study area of Hangzhou metropolis** (by Feng, 2002)

For the fractal distributions of cities, the calculation can be implemented from local parameters to global parameters. The local parameters, $\alpha(q)$ and $f(\alpha)$, can be directly computed by the $\mu$-weight method (Chhabra and Jensen,1989; Chhabra *et al*, 1989). However, the urban population density of Hangzhou is not of self-similar distribution. In this case, the $\mu$-weight method cannot be applied to the datasets displayed in Table 2. An advisable approach is to calculate the global parameter, $D_q$ and $\tau_q$, and then use Legendre transform to estimate the local parameters. Discretize equation (9) yields

$$\alpha(q)=\frac{\Delta\tau(q)}{\Delta q}=D_q+(q-1)\frac{\Delta D_q}{\Delta q}, \tag{13}$$

where Δ represents difference operator. Accordingly, equation (10) can be expressed as a discrete form

$$f(\alpha)=q\frac{\Delta\tau(q)}{\Delta q}-\tau(q)=q\frac{\Delta\tau(q)}{\Delta q}-(q-1)D_q. \tag{14}$$

Utilizing equation (13) and (14), we can estimate the local parameters $\alpha(q)$ and $f(\alpha)$ based on the global parameters. The smaller the $\Delta q$ value is, the exacter the $\alpha(q)$ and $f(\alpha)$ values will be.

**Table 2 The urban population density and the corresponding probability distribution of Hangzhou city (1964-2000)**

| Radius | Average population density ($\rho$) | | | | Probability distribution of population ($P$) | | | |
|---|---|---|---|---|---|---|---|---|
| ($r$) | 1964 | 1982 | 1990 | 2000 | 1964 | 1982 | 1990 | 2000 |
| 0.3 | 24130.876 | 29539.752 | 29927.903 | 28183.726 | 0.1966 | 0.1992 | 0.1699 | 0.1289 |
| 0.9 | 18965.755 | 22225.009 | 26634.162 | 26820.717 | 0.1545 | 0.1499 | 0.1512 | 0.1226 |
| 1.5 | 16281.905 | 18956.956 | 22261.980 | 24620.991 | 0.1326 | 0.1279 | 0.1264 | 0.1126 |
| 2.1 | 16006.650 | 19232.148 | 21611.817 | 23176.394 | 0.1304 | 0.1297 | 0.1227 | 0.1060 |
| 2.7 | 13052.016 | 15439.141 | 17290.295 | 18909.733 | 0.1063 | 0.1041 | 0.0982 | 0.0865 |
| 3.3 | 8259.322 | 9920.236 | 13178.503 | 19600.961 | 0.0673 | 0.0669 | 0.0748 | 0.0896 |
| 3.9 | 5798.447 | 7025.973 | 10537.808 | 16945.193 | 0.0472 | 0.0474 | 0.0598 | 0.0775 |
| 4.5 | 2625.945 | 3460.688 | 5559.761 | 10829.321 | 0.0214 | 0.0233 | 0.0316 | 0.0495 |
| 5.1 | 2142.703 | 2807.245 | 4180.368 | 7282.387 | 0.0175 | 0.0189 | 0.0237 | 0.0333 |
| 5.7 | 2141.647 | 2688.650 | 3923.003 | 6199.832 | 0.0174 | 0.0181 | 0.0223 | 0.0283 |
| 6.3 | 2185.160 | 2566.408 | 3515.837 | 5644.371 | 0.0178 | 0.0173 | 0.0200 | 0.0258 |
| 6.9 | 1438.027 | 1692.767 | 2197.220 | 4297.363 | 0.0117 | 0.0114 | 0.0125 | 0.0196 |
| 7.5 | 1083.473 | 1371.370 | 1795.763 | 3806.092 | 0.0088 | 0.0092 | 0.0102 | 0.0174 |
| 8.1 | 967.470 | 1256.167 | 1633.675 | 3152.766 | 0.0079 | 0.0085 | 0.0093 | 0.0144 |
| 8.7 | 842.494 | 1114.351 | 1442.105 | 2683.454 | 0.0069 | 0.0075 | 0.0082 | 0.0123 |
| 9.3 | 847.713 | 972.801 | 1265.412 | 2354.300 | 0.0069 | 0.0066 | 0.0072 | 0.0108 |
| 9.9 | 817.662 | 1050.963 | 1163.341 | 2028.299 | 0.0067 | 0.0071 | 0.0066 | 0.0093 |
| 10.5 | 812.050 | 1050.953 | 1143.197 | 1827.775 | 0.0066 | 0.0071 | 0.0065 | 0.0084 |
| 11.1 | 807.251 | 1050.998 | 1160.184 | 1651.076 | 0.0066 | 0.0071 | 0.0066 | 0.0075 |
| 11.7 | 625.112 | 979.407 | 1092.903 | 1580.848 | 0.0051 | 0.0066 | 0.0062 | 0.0072 |
| 12.3 | 691.323 | 901.339 | 1006.045 | 1490.260 | 0.0056 | 0.0061 | 0.0057 | 0.0068 |
| 12.9 | 574.569 | 870.020 | 972.123 | 1465.000 | 0.0047 | 0.0059 | 0.0055 | 0.0067 |
| 13.5 | 532.355 | 665.846 | 816.501 | 1278.000 | 0.0043 | 0.0045 | 0.0046 | 0.0058 |
| 14.1 | 381.306 | 486.856 | 679.057 | 1033.000 | 0.0031 | 0.0033 | 0.0039 | 0.0047 |
| 14.7 | 369.036 | 489.208 | 581.566 | 958.000 | 0.0030 | 0.0033 | 0.0033 | 0.0044 |
| 15.3 | 375.204 | 456.473 | 563.203 | 882.000 | 0.0031 | 0.0031 | 0.0032 | 0.0040 |

| Sum | 122755.471 | 148271.725 | 176133.732 | 218701.859 | 1.0000 | 1.0000 | 1.0000 | 1.0000 |
|---|---|---|---|---|---|---|---|---|

**Note**: The original data comes from Feng (2002). The probability data is calculated using the density data.

## 3.2 Results and findings

Now, a spatial analysis of Hangzhou's urban form can be made by means of the indexes based on the generalized multifractal spectrums. Using the models shown in Subsection 2.2 and the formulae given in Subsections 3.1, we can calculate the dummy correlation dimension $D_q$, mass exponent $\tau_q$, singularity exponent $\alpha(q)$, and local dimension $f(\alpha)$ for the four years, i.e., 1964, 1982, 1990, and 2000 (Table 3). The main parameter spectrum associated with spatial Renyi entropy is the generalized global dimension. In the dummy dimension spectrum, there are three common parameters, that is, capacity dimension $D_0$ ($q$=0), information dimension $D_1$ ($q$=1), and correlation dimension $D_2$ ($q$=2). Among these parameters, the most basic one is the capacity dimension. In the conventional multifractal spectrums, the capacity dimension is a fraction coming between the topological dimension and the Euclidean dimension of the embedding space ($0<D_0<2$). However, in this dummy multifractal spectrums, the capacity dimension $D_0$=2, which indicates a Euclidean dimension. For the distributions with characteristic scales such as exponential distribution and normal distribution, the capacity dimension proved to $d$=2 (Chen and Feng, 2012). Accordingly, the local fractal dimension $f(\alpha(0))=D_0=2$. Actually, the capacity dimension implies the maximum local dimension. In this sense, the conventional multifractal parameters are canonical multifractal parameters, while the generalized multifractal parameters are just dummy multifractal parameters.

**Table 3 Partial index values based on the generalized multifractal parameters of Hangzhou's urban population density distribution (1964-2000)**

| **Moment order** | | **1964** | | | | **1982** | | |
|---|---|---|---|---|---|---|---|---|
| ***q*** | $D_q$ | $\tau_q$ | $\alpha(q)$ | $f(\alpha)$ | $D_q$ | $\tau_q$ | $\alpha(q)$ | $f(\alpha)$ |
| -∞ | 3.5364 | -434.0879 | 3.5635 | 0.2376 | 3.5213 | -433.9992 | 3.5501 | -0.0038 |
| **-20** | 3.4185 | -71.7885 | 3.5635 | 0.5182 | 3.3934 | -71.2615 | 3.5427 | 0.4079 |
| **-15** | 3.3755 | -54.0073 | 3.5675 | 0.4952 | 3.3494 | -53.5906 | 3.5440 | 0.4305 |
| **-10** | 3.2943 | -36.2369 | 3.5755 | 0.4816 | 3.2675 | -35.9424 | 3.5506 | 0.4367 |
| **-5** | 3.0914 | -18.5485 | 3.5738 | 0.6796 | 3.0627 | -18.3762 | 3.5505 | 0.6237 |
| **-4** | 3.0110 | -15.0551 | 3.5631 | 0.8027 | 2.9814 | -14.9070 | 3.5363 | 0.7617 |
| **-3** | 2.9006 | -11.6024 | 3.5480 | 0.9585 | 2.8704 | -11.4817 | 3.5123 | 0.9449 |
| **-2** | 2.7388 | -8.2163 | 3.5354 | 1.1456 | 2.7100 | -8.1299 | 3.4846 | 1.1607 |
| **-1** | 2.4732 | -4.9465 | 3.4197 | 1.5268 | 2.4517 | -4.9035 | 3.3552 | 1.5483 |
| **0** | 2.0000 | -2.0000 | 2.4904 | 2.0000 | 2.0000 | -2.0000 | 2.4750 | 2.0000 |
| **1** | 1.5096 | 0.0000 | 1.5096 | 1.5096 | 1.5250 | 0.0000 | 1.5250 | 1.5250 |
| **2** | 1.3161 | 1.3161 | 1.2421 | 1.1681 | 1.3270 | 1.3270 | 1.2485 | 1.1699 |
| **3** | 1.2421 | 2.4842 | 1.1612 | 0.9995 | 1.2485 | 2.4969 | 1.1609 | 0.9859 |
| **4** | 1.2017 | 3.6050 | 1.1192 | 0.8718 | 1.2047 | 3.6141 | 1.1143 | 0.8433 |
| **5** | 1.1742 | 4.6967 | 1.0908 | 0.7574 | 1.1746 | 4.6983 | 1.0827 | 0.7153 |
| **10** | 1.1013 | 9.9116 | 1.0248 | 0.3366 | 1.0948 | 9.8532 | 1.0132 | 0.2790 |
| **15** | 1.0685 | 14.9591 | 1.0074 | 0.1513 | 1.0603 | 14.8448 | 0.9976 | 0.1186 |
| **20** | 1.0508 | 19.9658 | 1.0023 | 0.0802 | 1.0423 | 19.8045 | 0.9935 | 0.0654 |
| ∞ | 1.0009 | 434.6232 | 0.9986 | 0.0006 | 0.9926 | 435.0063 | 0.9903 | 0.0006 |

Continued from Table 3

| **Moment order** | **1990** | | | | **2000** | | | |
|---|---|---|---|---|---|---|---|---|
| ***q*** | $D_q$ | $\tau_q$ | $\alpha(q)$ | $f(\alpha)$ | $D_q$ | $\tau_q$ | $\alpha(q)$ | $f(\alpha)$ |
| -∞ | 3.4985 | -434.6922 | 3.5265 | 0.0506 | 3.3582 | -434.0431 | 3.3844 | -0.0063 |
| **-20** | 3.3717 | -70.8057 | 3.5254 | 0.2985 | 3.2294 | -67.8164 | 3.3797 | 0.2234 |
| **-15** | 3.3264 | -53.2218 | 3.5268 | 0.3198 | 3.1854 | -50.9665 | 3.3774 | 0.3052 |
| **-10** | 3.2425 | -35.6671 | 3.5290 | 0.3773 | 3.1058 | -34.1639 | 3.3745 | 0.4193 |
| **-5** | 3.0397 | -18.2382 | 3.5105 | 0.6857 | 2.9171 | -17.5028 | 3.3554 | 0.7257 |
| **-4** | 2.9612 | -14.8061 | 3.4958 | 0.8228 | 2.8441 | -14.2205 | 3.3452 | 0.8398 |
| **-3** | 2.8543 | -11.4172 | 3.4796 | 0.9784 | 2.7439 | -10.9755 | 3.3319 | 0.9798 |
| **-2** | 2.6980 | -8.0939 | 3.4669 | 1.1602 | 2.5969 | -7.7906 | 3.3011 | 1.1884 |
| **-1** | 2.4417 | -4.8834 | 3.3250 | 1.5583 | 2.3621 | -4.7243 | 3.0864 | 1.6379 |
| **0** | 2.0000 | -2.0000 | 2.4355 | 2.0000 | 2.0000 | -2.0000 | 2.3272 | 2.0000 |
| **1** | 1.5645 | 0.0000 | 1.5645 | 1.5645 | 1.6728 | 0.0000 | 1.6728 | 1.6728 |
| **2** | 1.3847 | 1.3847 | 1.3128 | 1.2409 | 1.5226 | 1.5226 | 1.4578 | 1.3930 |
| **3** | 1.3128 | 2.6257 | 1.2351 | 1.0797 | 1.4578 | 2.9157 | 1.3884 | 1.2495 |
| **4** | 1.2740 | 3.8219 | 1.1974 | 0.9677 | 1.4231 | 4.2694 | 1.3565 | 1.1567 |
| **5** | 1.2485 | 4.9938 | 1.1739 | 0.8756 | 1.4009 | 5.6037 | 1.3374 | 1.0834 |
| **10** | 1.1857 | 10.6709 | 1.1220 | 0.5488 | 1.3482 | 12.1339 | 1.2954 | 0.8203 |
| **15** | 1.1580 | 16.2118 | 1.1045 | 0.3560 | 1.3252 | 18.5526 | 1.2802 | 0.6505 |
| **20** | 1.1422 | 21.7014 | 1.0970 | 0.2390 | 1.3118 | 24.9234 | 1.2727 | 0.5308 |
| ∞ | 1.0908 | 434.9397 | 1.0880 | 0.0007 | 1.2614 | 434.8733 | 1.2578 | 0.0009 |

**Note**: For 1964 and 2000 years, if $q\to-\infty$, then $f(\alpha)<0$. These are abnormal results. This suggests that if the order of moment becomes too small, the spectrums of local fractal dimension may not converge.

The spatial information content reflected by individual value of a parameter is limited. However, comparing a set of values of a parameter under different conditions in given time or the single parameter's values in different times, we can obtain useful information about spatial and temporal evolution of a city. Where Hangzhou is concerned, urban population density, on the whole, followed

Clark's law (Clark, 1951). We can use the exponential function or generalized exponential function to model the population density decay (Feng, 2002). From 1964 to 1982 to 1990, the urban density of central area went up and up, but from 1990 to 2000, the central density went down (Table 2). The process of Hangzhou's urban growth can be reflected by the global parameters spectrums (Figure 2). The dummy dimension spectrum is a sigmoid decay curve. From 1964 to 2000, the value of the maximum dummy correlation dimension $D_{-\infty}$ went down, but the value of the minimum dummy correlation dimension $D_{-\infty}$ went up. In short, the range of general correlation dimension became shorter and shorter (Figure 2(a)). An index of half range of general correlation dimension can be defined as

$$\gamma = \frac{D_{-\infty} - D_{\infty}}{2}, \tag{15}$$

which can be used to measure the decrease of population density difference. For the four years, the half range values are 1.2677, 1.2643, 1.2039, and 1.0484 (Table 4). From 1964 to 1982, the city grew slowly because of the political movement of Great Cultural Revolution (1966-1976). However, from 1982 to 2000, the city grew faster and faster. The half range mirrors the rate of urban growth. A mass exponent spectrum comprises two straight line segments. The change of urban form can be reflected by the difference of the slopes of the two line segments on the plot (Figure 2(b)). From equation (7) it follows

$$\frac{\tau_q}{q} = D_q - \frac{D_q}{q}, \tag{16}$$

which is an approximation formula of the slopes of the mass exponent against the moment order. According to the property of the $D_q$ function, when $q \to \pm\infty$, $D_q/q \to 0$. Therefore, if $q \to -\infty$, the slope $\tau_q/q \to D_{-\infty}$; if $q \to \infty$, the slope $\tau_q/q \to D_{\infty}$. This implies that the difference between the two slopes of the mass exponent curve is theoretically equal to the range of the dummy fractal dimension. The density gradient change of urban population distribution from 1964 to 2000 can be described with the slope differences: 2.5355, 2.5287, 2.4078, and 2.0968 (Table 4).

The evolution of Hangzhou's urban population can also be reflected by the local parameter spectrums. The relationship between the singularity exponent and the corresponding local fractal dimension form a unimodal curve, which is termed $f(\alpha)$ curve and represents the common multifractal spectrum (Feder, 1988). In our context, of course, we have dummy multifractal spectrum of local structure. Owing to the density gradient went down and down, the span of the $f(\alpha)$ curve becomes smaller and smaller, and the weights of the left and right sides reverse (Figure 3). The left end corresponds to the positive order of moment ($q$>0, $q$→∞), while the right end corresponds to the negative order of moment ($q$<0, $q$→-∞). The former indicates the center and sub-centers of Hangzhou with higher density, while the latter indicates the peripheral regions of the city with lower density. The shape of the $f(\alpha)$ curves mirror the change of urban population density. In other words, the $f(\alpha)$ spectrums can provide us with an image of the spatio-temporal evolution of urban density. In light of the change trend of spectral curves, the difference between high density areas and low density areas became smaller and smaller from 1964 to 2000.

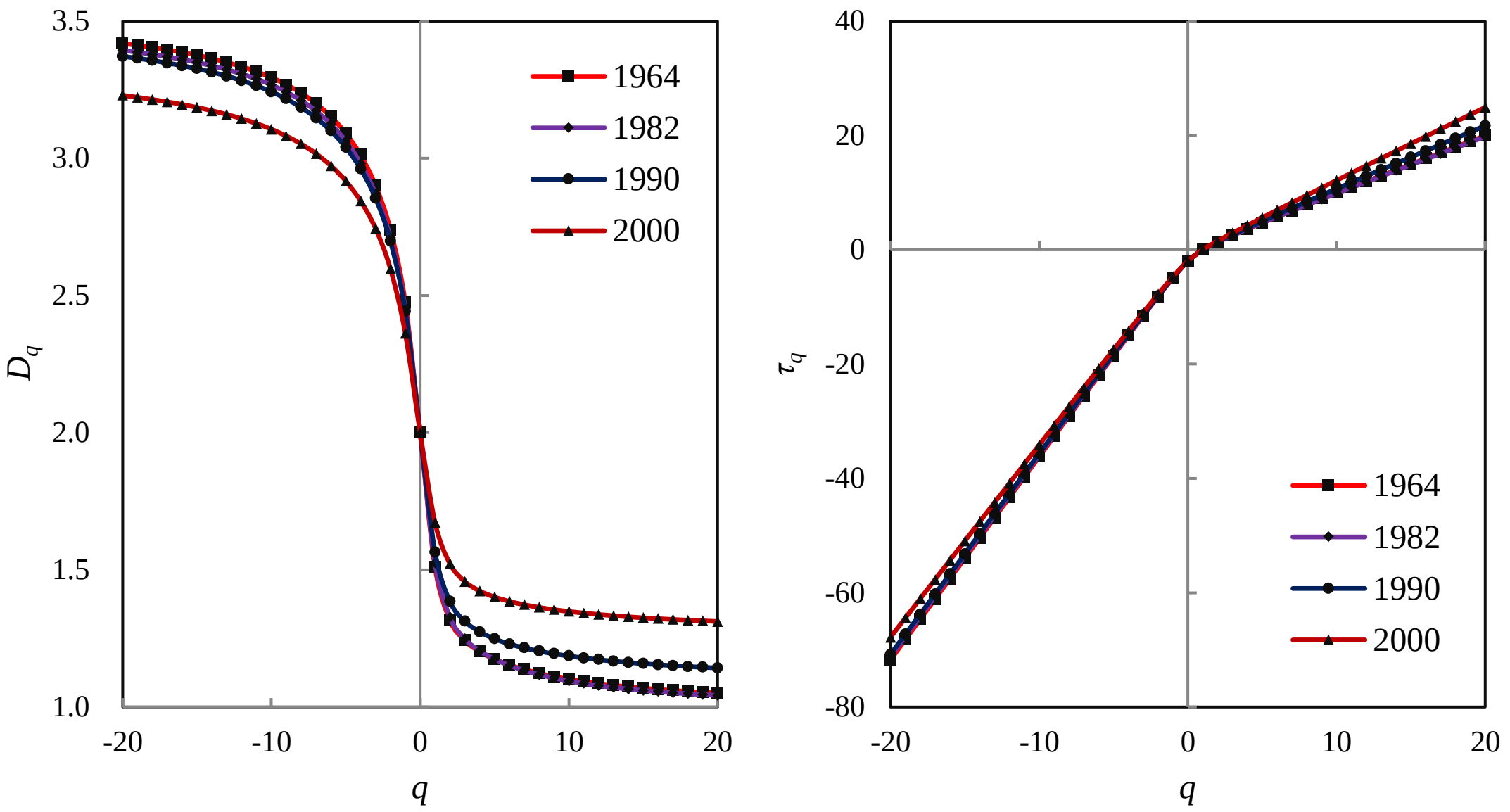

a. Dummy correlation dimension b. Dummy mass exponent

**Figure 2 The dummy correlation dimension spectrums and mass exponent curves of Hangzhou's urban population density distribution (1964-2000)**

Shannon's information entropy can be regarded as a special case of Renyi entropy. We can examine information entropy values and the related indexes of urban spatial form. Shannon entropy can be calculated with equation (12). The normalized entropy is the ratio of Shannon entropy to the maximum entropy, $M_{max}$=ln(26)≈3.2581 nat. As indicated above, doubling the normalized Shannon entropy yields the dummy information dimension, i.e., $D_1=2D_1^*$. The half range of the general correlation dimension can be computed by equation (15). The slope difference is the range of the slopes of the dummy mass exponent. Correspondingly, the density gradients can be evaluated by Clark's law (Clark, 1951). The reciprocals of the density gradients are the characteristic radius of urban population distribution. All these indexes are displayed in Table 4. Along with urban growth, the density gradient descended, thus the characteristic radius of urban population distribution extended. In fact, from 1964 to 2000, the density gradient of population distribution from core to periphery descended gradually: from 0.2805 to 0.2724 to 0.2756 to 0.2534, and accordingly, the characteristic radius of population distribution ascended gradually: from 3.5644 to 3.6713 to 3.6285 to 3.9457 km. As a result, the spatial information entropy and information dimension went up and up. The range of the dummy correlation dimension and the slope difference of mass exponent went down and down. The spatial indexes of urban form are correlated with one another, but there is difference between the indexes based on density gradient and the ones based on spatial information entropy. Combining these indexes and the dummy multifractal parameters, we can make spatial analysis of cities in more efficient way.

**Table 4 The spatial entropy and related parameters of Hangzhou's urban population density distribution (1964-2000)**

| Year | Density gradient | Characteristic radius | Shannon entropy | Normalized entropy | Information dimension | Half range | Slope difference | Areal entropy |
|---|---|---|---|---|---|---|---|---|
| **1964** | 0.2805 | 3.5644 | 2.4593 | 0.7548 | 1.5096 | 1.2677 | 2.5355 | 3.1943 |
| **1982** | 0.2724 | 3.6713 | 2.4843 | 0.7625 | 1.5250 | 1.2643 | 2.5287 | 3.1896 |
| **1990** | 0.2756 | 3.6285 | 2.5486 | 0.7822 | 1.5645 | 1.2039 | 2.4078 | 3.1944 |
| **2000** | 0.2534 | 3.9457 | 2.7251 | 0.8364 | 1.6728 | 1.0484 | 2.0968 | 3.1811 |

**Note**: The areal entropy is the spatial entropy of the cumulative distribution of urban population density. The unit of spatial entropy is *nat* because the formula is on the base of natural logarithm.

The spatial entropy based on the average density of urban population by means of rings is in fact a 1-dimensional spatial measurement. The areal entropy based on 2-dimensional space can be calculated through spatial cumulative distribution of urban population density. From 1964 to 2000, the areal entropy values are 3.1943, 3.1896, 3.1944, and 3.1811, bearing no significant change (Table 4). This suggests that the areal entropy defined in 2-dimensional space approaches to a constant and cannot reflect urban growth. It is the density entropy rather than the areal entropy that can be employed to reflect urban evolution. In contrast, the areal entropy seems to indicate some conservation law of urban evolution.

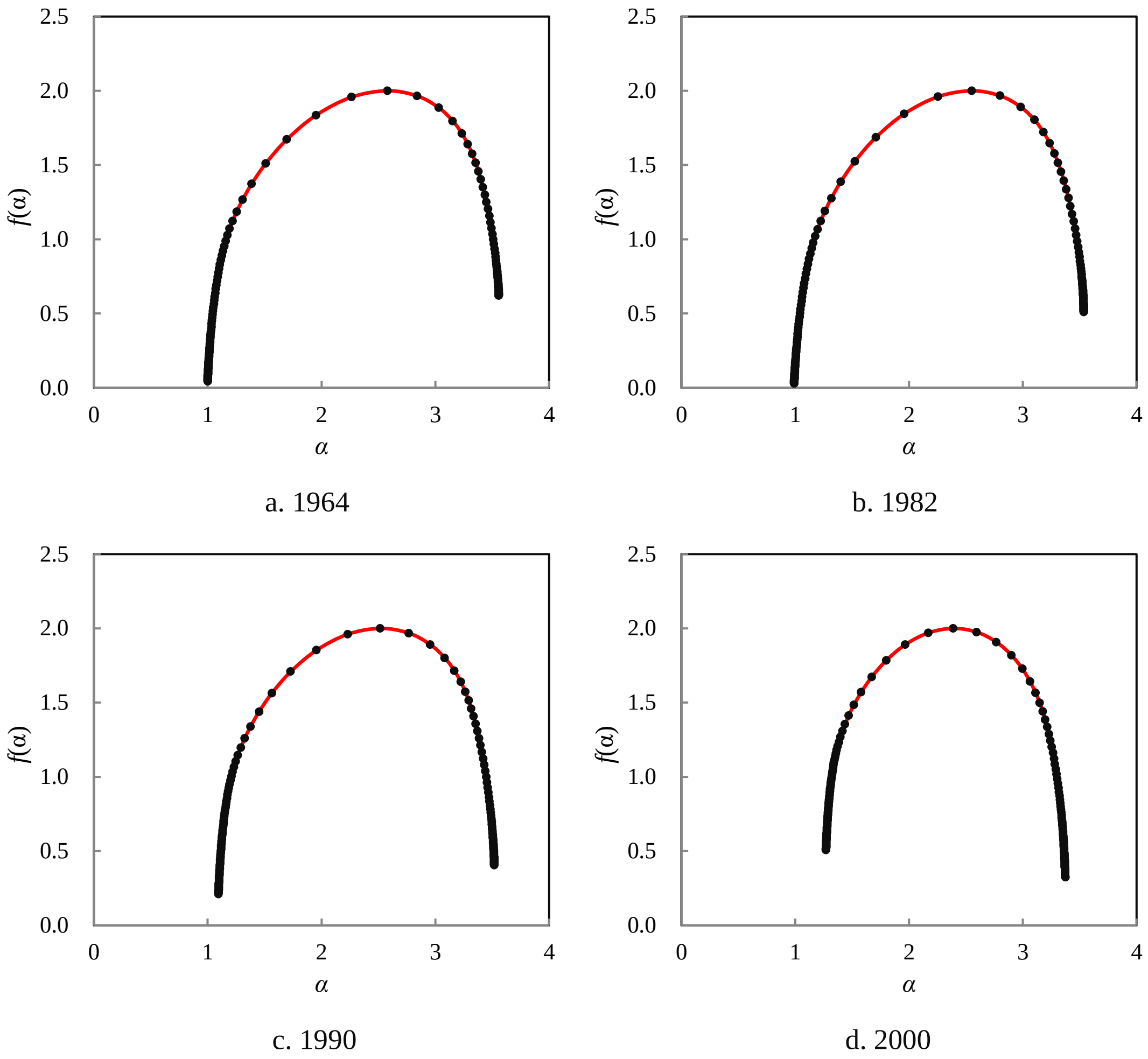

**Figure 3 The dummy singularity spectrums of the spatial structure of Hangzhou's population density distributions**

Spatial entropy and fractal dimension are two measurements of space filling and distribution uniformity. Also, they are measures of spatial complexity. Now, the principal conclusions about Hangzhou's urban population distribution and evolution can be reached as follows. **First, Hangzhou's urban population distribution became more and more uniform. Meantime, the urban space was more and more occupied by buildings.** This feature can be found by information entropy values. Spatial entropy increase indicates that the density decay rate from the urban core to the periphery became smaller and smaller from 1964 to 2000. A process of population suburbanization might take place at the turn of the century (Chen and Jiang, 2009; Feng, 2002; Feng and Zhou, 2005). **Second, Hangzhou's urban growth takes on an accelerated process.** This property can be recognized by the dummy multifractal parameters. The difference between the dummy multifractal dimension spectrum curves for 1964 and that for 1982 is not clear, but the difference between the spectrum curve for 1982 and that for 1990 is significant. Especially, the difference between dummy multifractal dimension spectrum curve for 1990 and that for 2000 is more significant. The spectrum curve difference between two immediate census years become clearer and clearer from 1964 to 2000 (Figures 2 and 3). **Third, Hangzhou's urban pattern became more complex, but the structure was improved.** Along with urban growth, the dummy multifractal parameter spectrums approached to reasonable intervals. The reasonable range of the dummy generalized multifractal dimension is from 1 to 3. From 1964 to 2000, the $D_{-\infty}$ value became closer to 3, while the $D_{\infty}$ value ascended from about 0.99 to about 1.26 (Table 3). Urban evolution is a process of spatial optimization by self-organization, but the precondition of self-organized optimization is rational policy of city development and management.

## 4. Discussion

This is a methodological study on generalization and application of multifractal description and analysis. The monofractal method is based on simple fractals, and can be only applied to fractal objectives. A fractal phenomenon is free of characteristic scales, and cannot be described with the traditional mathematical method based on the concept of characteristic scale. Fractal geometry is an effective approach to modeling scale-free phenomena such as urban form and growth. However, multifractal scaling is different from monofractal scaling. Multifractal theory provides a quantitative tool for describing a broad range of heterogeneous phenomena, and there is an analogy between multifractal scaling and thermodynamic entropy (Stanley and Meakin, 1988). Formally, multifractal parameters can be generalized to characterize both fractal patterns and non-fractal objects. The generalized results are termed dummy multifractal parameters and spectrums. The theoretical base of generalizing multifractal measures lies in the following relation: the normalized entropy is equal to the normalized fractal dimension (Chen, 2017). The key principle rests with the dual relation between Euclidean geometry and fractal geometry. For Euclidean geometry, the dimension is known and possesses no spatial information, and we need the conventional measures such as length, area, and volume; in contrast, for fractal geometry, the conventional measures are known and give little spatial information, and we should make use of the dimension. The complementary set of a fractal set is a Euclidean object. Both the fractal set and its complement can be expressed with hierarchical scaling, and both the Euclidean geometry and fractal geometry are involved with a power function. For multifractal structure, we have

$$N_0(\varepsilon) \propto \varepsilon^{-f(\alpha(0))}, \qquad (17)$$

in which the zero order local dimension $f(\alpha(0)) = D_0 < 2$. For Euclidean structure, however, the parameter $f(\alpha(0)) = D_0 = 2$, thus equation (17) changes to

$$N_0(\varepsilon) \propto \varepsilon^{-2}, \qquad (18)$$

which is suitable for exponential distribution (Chen and Feng, 2012). The conventional multifractal models are based on equation (17), while the dummy multifractal models are based on equation (18). The above empirical analysis lends support to the theoretical suggestion that the dummy multifractal parameters can be used to describe non-fractal distribution such as urban population density. The calculation process relies heavily on spatial Renyi entropy. In the canonical multifractal spectrums, capacity dimension is a fractal dimension (fractional dimension); while in the dummy multifractal spectrums, capacity dimension is a Euclidean dimension (integral dimension). The values of capacity dimension make a criterion, by which we can distinguish the conventional multifractal spectrum from dummy multifractal spectrums (Table 5).

**Table 5 Comparison between multifractal parameters and generalized multifractal parameters**

| Item | Canonical multifractal parameters | Dummy multifractal parameters |
|---|---|---|
| **Basis** | Renyi entropy | Renyi entropy |
| **Key relation** | $N(\varepsilon)=\varepsilon^{-f(\alpha(0))}$ | $N(\varepsilon)=\varepsilon^{-2}$ |
| **Objective** | Scale-free distributions | Distributions with characteristic scales |
| **Capacity dimension** | Fractal dimension ($0 \le D_0 \le 3$) | Euclidean dimension ($D_0$=1, 2, or 3) |
| **General correlation dimension value** | Fall into the range from topological dimension to the Euclidean dimension of embedding space ($0 \le D_q \le 2$ or 3) | Go out of the range from topological dimension to the Euclidean dimension of embedding space ($0 \le D_q \le 4$ or 5) |
| **Calculation process** | The best approach is from local parameters to global parameters, the alternative approach is from global parameters to local parameters | The only approach is from global parameters to local parameters for the time being. No other approach has been found so far. |

Entropy, especially spatial Renyi entropy, is an important spatial measurement of urban and regional systems. Renyi entropy has been applied to measuring regional land use and urban sprawl (Fan *et al*, 2017; Padmanaban *et al*, 2017), and the interesting results show that the entropy measurement effect is similar to the fractal dimension. Just based on the Renyi entropy, the general correlation dimension of multifractals is defined to describe the scaling phenomena (Feder, 1988). Multifractal geometry becomes a powerful tool for multi-scaling analysis of scale-free systems. It has been employed to characterize city-size distributions (Chen and Zhou, 2004; Haag, 1994), urban form and growth (Ariza-Villaverde *et al*, 2013; Chen and Wang, 2013; Frankhauser, 1998; Murcio *et al*, 2015), urban residential land price (Hu *et al*, 2012), regional population distributions (Appleby, 1996; Liu and Liu, 1993; Sémécurbe *et al*, 2016), central place networks (Chen, 2014), bus-transport network (Pavón-Domínguez *et al*, 2017), and so on. In the previous works, the multifractal method has been generalized for two times. One is generalized to model Zipf's distribution of city sizes (Chen and Zhou, 2014), and the other is to define urban-rural territory structure (Chen, 2016). The generalized multifractal models have been built for the rank-size distribution of cities and urbanization process and patterns. In this paper, the multifractal modeling is generalized to describe spatial distribution with characteristic scales. Urban population density distribution follows Clark's law, which takes on negative exponential distribution (Feng, 2002; Clark, 1951). Exponential distribution bears a characteristic length indicating the average radius of human activities (Takayasu, 1990). This kind of distribution cannot be depicted by the common fractal method, but maybe it bears hidden self-affine fractal pattern (Chen and Feng, 2012). However, based on Renyi entropy, urban population density can be modeled with the spatial index sets based on above-developed dummy multifractal spectrums.

As a matter of fact, the dummy multifractal parameters can be estimated by means of the spatial datasets based on arbitrary zonal systems. The statistical average based on concentric circles on digital maps is not a requirement. The merits of changing the zonal data into the ring data are as follows. On the one hand, the negative influence of the size and shape differences of zones on spatial analysis can be lessened to some extent. A zonal system comprises small regions of varied sizes and shapes. Data smoothing can reduce the random disturbance of spatial irregularity. On the other, more importantly, it is easy to judge the property of urban population density distribution. If it bears no characteristic scales, we can use the conventional multifractal measurements, otherwise, we can use the dummy multifractal indexes. Though the multifractal method is advanced efficiently, the deficiencies of this study are inevitable. The main shortcomings are as below. First, only urban population density distribution is testified. In theory, the method can be applied to other spatial distributions without characteristic scales, but no more case studies are presented owing to the limit of space of a paper. Second, the calculation is based on the process from global parameters to local parameters, no more convenient approach is proposed. Third, the observational datasets of Hangzhou city are not new. Despite the fact that new or old data have no significant influence on methodological studies, it is better to find newer data for an empirical analysis. It is impossible to solve too many problems in one work. The pending questions remain to be answered in future studies.

# 5. Conclusions

As indicated above, spatial analysis of cities fall into two groups: one is the spatial modeling based on characteristic scales, and the other is the spatial description based on sets of scaling exponents. This paper is devoted to finding the links between the traditional scale-based spatial analysis on characteristic scales and future scaling-based spatial analysis of cities. From the theoretical models and empirical case, the main conclusions can be drawn as follows.

**First, the multifractal measures can be generalized to describe the geographical distributions without characteristic scales.** The conventional spatial modeling is based on characteristic scales such as typical distance, average values, standard deviations, and eigenvalues. Fractal geometry is mainly applied to the scale-free distributions. Spatial Renyi entropy is an important concept which can be used to associate the characteristic distributions and scaling distributions. The conventional spatial entropy is measured by way of geographical zonal systems and depends on the scales of spatial measurement. Based on the relation between the normalized spatial Renyi entropy and normalized fractal dimension, a new parameter bearing an analogy with the generalized correlation dimension of multifractal theory can be defined and termed dummy correlation dimension. As a result, we can derive a set of parameters including dummy mass exponent, dummy singularity exponent and dummy local fractal dimension. The dummy multifractal parameters are not real fractal parameters. They are just multi-scale indexes of spatial structure bearing an analogy with multifractal parameters in fractal geometry.

**Second, the dummy multifractal parameters can be estimated by using normalized spatial**

**Renyi entropy.** The global parameters of multifractals are defined on the base of Renyi entropy. In this paper, the dummy multifractal parameters rely much heavily on the spatial Renyi entropy. The dummy multifractal parameters can be calculated as long as the Renyi entropy is computed. The process of measurement, calculation, and analysis is as below: (1) By means of zonal systems, grids, or systems of concentric circles, we can convert a dataset of spatial distribution into a dataset of probability distribution. (2) Based on probability distribution, spatial Renyi entropy can be calculated. (3) Normalizing spatial Renyi entropy, we can work out a set of dummy multifractal parameters. (4) Finally, using the dummy multifractal parameter spectrums, we can make spatial analysis of city development. Though the generalized multifractal parameters are not real multifractal parameters, they are applicable to urban spatial analysis in practice.

**Third, the sphere of application of the dummy multifractal parameters is much larger than that of the canonical multifractal parameters.** In theory, any spatial distributions can be converted into probability distributions, and thus can be described with spatial Renyi entropy. The canonical multifractal geometry is only applicable to power-law distribution. Since the dummy multifractal parameters can be applied to exponential distribution, it can be applied to normal distribution, lognormal distribution, Poisson distribution, Weibull distribution, and so on. Differing from the normative multifractals with capacity dimension coming between the topological dimension and the Euclidean dimension of the embedding space, the dummy multifractal parameter spectrums bear an integral value of capacity dimension. Therefore, in light of the capacity dimension value, we can distinguish the canonical multifractal parameter spectrums from the dummy multifractal parameter spectrums.

## Acknowledgements

This research was sponsored by the National Natural Science Foundations of China (Grant No. 41671167 & 41671157). The supports are gratefully acknowledged. We are also grateful to two anonymous reviewers whose constructive suggestions were very helpful in improving the paper’s quality.